\documentstyle{mn}
\input{epsf}

\title[Scaling in Gravitational Clustering]
{Non-local scaling in two--dimensional gravitational clustering} 

\author[Munshi et al.]{Dipak Munshi$^{1}$, Lung-Yih
Chiang$^{1}$, Peter Coles$^{1,2}$ and Adrian L. Melott$^{2}$ \\
$^{1}$ Astronomy Unit, Queen Mary \& Westfield College, 
University of London, London E1 4NS, UK\\
\smallskip
$^{2}$ Department of Physics \& Astronomy, University of Kansas, Lawrence,
Kansas 66045, USA.}

\pagerange{000,000}
\pubyear{1997}

\begin{document}

\maketitle

\begin{abstract}
Using an ensemble of high resolution 2D numerical simulations, we explore the
scaling properties of cosmological density fluctuations in the
non-linear regime. We study the scaling behaviour of the usual
$N$--point volume-averaged correlations, but also examine the scaling of the
entire probability density function (pdf) of the fluctuations.
We focus on two important issues: (i) whether
the scaling behaviour of 2D clustering is consistent with what one infer
from radial collapse arguments; and (ii) whether there is any evidence from 
these high-resolution simulations that a regime of {\em stable clustering}
is ever entered. We find that the answers are (i) yes and (ii) no.
We further find that the behaviour of the highly non-linear
regime in these simulations suggests the
existence of a regime where the correlation function is
{\em independent} of the initial power spectrum.
\end{abstract}


\section{Introduction}
The standard model of cosmological structure formation is based on the idea
that small initial density fluctuations grew by the action of gravitational
instability into the large inhomogeneities we observe today. While the
initial, linear stages of fluctuation growth have been understood analytically
for many years, the later stages are less amenable to analytical study because
these are described by non-linear gravitational physics 
involved (e.g. Sahni \& Coles 1995). The standard approach for the early
stages is via perturbation theory, where an expansion in an appropriately
chosen small parameter is employed (e.g. Fry 1984; Moutarde et al. 1991; Buchert
1992; Bernardeau 1992; Munshi et al. 1997). Such an expansion is typically taken to first order
when the density fluctuations are very small in amplitude, but some notable
successes have been achieved in the weakly non-linear regime when
the density fluctuations are of order unity, 
by taking perturbation theory to higher orders. On the other hand,
the strongly non-linear regime where the density fluctuations exceed
unity by a large factor is intrinsically non-linear and a conceptually
different approach is required. One method involves adopting the
so-called {\em stable clustering} ansatz (e.g. Peebles 1980), but this
ansatz only applies (if at all) when density contrasts are very large indeed
(Jain 1997). Probing the regime leading up to the stable clustering limit is considerably
more difficult, but some progress has been made using arguments based
on the properties of spherical collapse (Hamilton et al. 1991;
Nityananda \& Padmanabhan 1994; Jain, Mo \& White 1995; 
Padmanabhan 1996; Padmanabhan et al. 1996; Peacock \& Dodds 1996).
These arguments allow a simple scaling model to be constructed, which
can describe certain aspects of  the statistical behaviour of density fluctuations
all the way from the linear regime to the stable clustering limit by 
a kind of functional interpolation between the linear and stable clustering 
limits,
guided by the behaviour of numerical simulations. Since the original paper
by Hamilton et al. (1991), there has been an exploration of the 
physical origin of this functional behaviour
(Nityanada \& Padmanabhan 1994; Padmanabhan 1996; 
Padmanabhan et al. 1996; Munshi \& Padmanabhan 1997), and some generalisations
of the original arguments have been presented (e.g. Jain, Mo \& White 1995;
Peacock \& Dodds 1996).

In this paper we address two issues related to these scaling arguments, and the
models emerging from them. The first
concerns the intermediate regime. 
We argue that the behaviour
of correlation functions in this regime is basically governed by the
number of spatial dimensions. We confirm this argument using a battery
of two-dimensional $N$-body experiments, and find the results to be
consistent with our assertion, lending theoretical understanding to
the numerical fits obtained from three-dimensional simulations. An
additional advantage of using two-dimensional simulations is that it is 
possible to obtain high resolution at relatively small computational cost 
compared to full three-dimensional simulations. With this in mind we also 
address the (related) issue of the stable clustering limit itself, 
and whether there is any evidence from our simulations that this description 
is appropriate at all during the highly non-linear stages.

\section{Scaling in Gravitational Clustering}

The evolution of the volume-averaged two-point correlation 
function $\bar \xi_2(x,a)$ in a 
D-dimensional space can be described by the following equation:
\begin{equation}
{\partial \Xi \over \partial  A} - h{\partial \Xi \over \partial  X} = Dh
\label{twentynine}
\end{equation}
(Peebles 1980),  where we have introduced the following  
variables to replace the comoving 
co-ordinate $x$ and scale factor $a$:
\begin{equation}
 \Xi = \ln[(1 + \bar \xi_2(x,a)], ~~~~ A = \ln a, ~~~~  X = \ln x\,.
\end{equation}
The quantity $h$ is defined through the relation 
$ v = -h \dot a x $where $v$ is the mean pair velocity at separation x.
The characteristic of the same equation represents conservation of pairs
during collapse:
\begin{equation}
l^D = x^D(1 + \bar \xi_2(x,a));
\end{equation}
we have introduced a new length scale $l = \langle x^D \rangle^{1/D}$
which represents the typical initial (Lagrangian) length scale from which 
structures are collapsing at a certain epoch. 
This suggests that the nonlinear correlation
function can be expressed as a function of the initial (linear) 
correlation function, evaluated at a different length scale to that
at which the final function is given, i.e. $\xi_2(x,a) = F_n[\xi_2(l,a)]$, 
where the subscript $n$ indicates that the function $F$ is permitted
also to depend on the initial spectral index. Since in
scale-free gravitational clustering all the one point moments of the 
distribution
depend on epoch $a$ and length scale $l$ only through the volume averaged  
correlation 
function $\bar \xi_2(x,a)$,
evolution of any statistical property can be expressed in terms of such a 
non-local mapping.
For example, the probability distribution function (pdf),
$P_\delta(x,a)$,  of the density  field  ($\delta$) can be expressed 
as a function of initial pdf at length  scale $l$, $P_\delta (x,a) = 
{\cal F}_n[P_\delta(l,a)]$; $\delta$ acts as a
parameter in such a scaling.  Higher order 
moments of  $P_\delta$ (which are volume-averaged higher-order correlation 
functions) will also exhibit scaling properties due to the hierarchical
nature of the clustering although knowledge of all the higher order
correlation functions does not necessarily specify pdf uniquely
(cf. Coles \& Jones 1991)

Simulations suggest that gravitational
clustering evolves through three different phases. 
The early phase of
clustering, during which the fluctuations are small,
can be studied using perturbative methods. The evolution of 
higher order correlation functions in this epoch is dominated by precollapse  
evolution of rare events which can be described by a 
spherical collapse model  (Bernardeau 1992, 1994). 
Scaling takes a very simple form 
in this case, with  $\bar \xi_2(x,a) \propto [ \bar \xi_2(l,a)]$ where $x$ and 
$l$ are not very different. Consequently 
$h$ increases almost linearly with $\bar \xi_2(x,a)$ in this epoch.

The intermediate regime can be
characterized by turn-around and collapse of these very high density peaks,
leading to a scaling described by $\bar \xi_2(x,a) \propto [\bar \xi_2(l, a)]^{D}$. 
Hence in any dimension $D$, this relation  has a slope exactly equal to
the slope of background space. One of the motivation of this paper is to
check this prediction against N-body simulations in 2D. 
In 3D it has already been studied and shown to follow the predicted scaling 
(Munshi \& Padmanabhan 1997). In this regime, 
$h$ reaches a maximum value of $2$ at turn around  before 
decreasing in highly nonlinear regime when virialised structures start to 
collapse.

It is generally believed that collapsed haloes are no longer
participating in the global expansion and therefore that the
average peculiar pair velocity counterbalances the  Hubble expansion. 
This would mean that $h$ attains a 
asymptotic value of $1$ in the final regime of strong nonlinearity. 
Such reasoning leads to the  
``stable clustering'' {\em ansatz} which will produce a scaling 
$\bar \xi_2(x,a) \propto [ \bar \xi_2(l, a)]^{D/2}$. So if
stable clustering is a good approximation, the slope of the scaling function is
again determined only by dimension of the background space.

Scaling of the   higher-order correlations can be analysed by studying the
evolution of the so called $S_N$ parameters.  To summarize in general we have 
following forms of higher order correlation
functions in the {\em perturbative}, {\em intermediate} and {\em nonlinear}
regimes discussed above:
\begin{eqnarray}
\bar \xi_N^{pert}(x,a) &=& S_N^{tree} \bar \xi_{2,lin}^{N-1}(l,a); \nonumber \\
\bar \xi_N^{int}(x,a) &=& A^{N-1} S_N^{int} \bar \xi_{2,lin}^{D(N-1)} (l,a); \nonumber \\
\bar \xi_N^{non}(x,a) &=& B^{N-1} S_N^{non}  \bar \xi_{2,lin}^{D(N-1)/2} (l,a).  
\end{eqnarray}
In these equations $A$ and $B$ are constants determined from the scaling of 
two point correlation function;
the $S_N$ parameters are defined to be the dimensionless
ratios of higher order correlation functions with suitably raised powers
of two point correlation functions $ S_N = \xi_N(x,a)/ \xi_2(x,a)^{N - 1}$ 
at same scale, and by definition $S_1 = S_2 = 1$. For the purposes of
this paper we are only interested in the slopes of the inferred scaling
relations, so we use the quantities $S_N$  as floating parameters
in different regimes and use only the slopes in the following analysis.

Using these scaling relations (4),
it is easy to construct the complete dependence of the two-point 
correlation functions on 
spatial and temporal coordinate i.e. $x$ and $a$ in different regime.
\begin{equation}
\bar \xi_2(x,a) = \bar \xi_0 a^{2Dh/[ 2 + h(n+D)]} x^{-Dh(n+D)/[2 + h(n+D)]},
\end{equation}
where $n$ is the spectral index of initial scale free power spectrum.
Note that equation (5) is only a solution of equation (1) for 
$\bar \xi_2(x,a) \gg 1$, i.e. in the highly nonlinear regime.
This expression makes it clear that assumption of stable clustering
leads to a final correlation function which depends explicitly on 
initial power spectrum.  On the other hand one might expect that highly 
nonlinear gravitational physics might lead to correlation functions that
do not  depend explicitly on the initial power spectrum. This
 would translate into a behaviour of $h$ of the form  $h(n+D) = K$, where 
$K$ is a constant. Note that the mean pair velocity, however, is still
governed by a dependence on the initial power-spectrum in this case.
It is interesting to note that if stable clustering is a good
approximation, in 2D the slope of the scaling function in this limit should
be the same as it is in the linear regime.

Although several studies have been carried out to check stable clustering 
ansatz (e.g. Jain, Mo \& White 1995; Jain 1997), such studies are  affected 
by limited dynamical range available for 3D N-body simulations. Since 2D 
provides much wider dynamic range it is much easier to study  extreme nonlinear
 regime with very high values of $\bar \xi_2(x,a)$. It is also worth mentioning that
most of the earlier studies have been performed using 
$P^3M$ codes and it is not very clear if sub-grid resolution in such
simulations produce different results compared to more robust 
$PM$ simulations (Splinter et al. 1997). 

\begin{figure*}
\protect\centerline{
\epsfysize = 7.5truein
\epsfbox[20 136 587 764]
{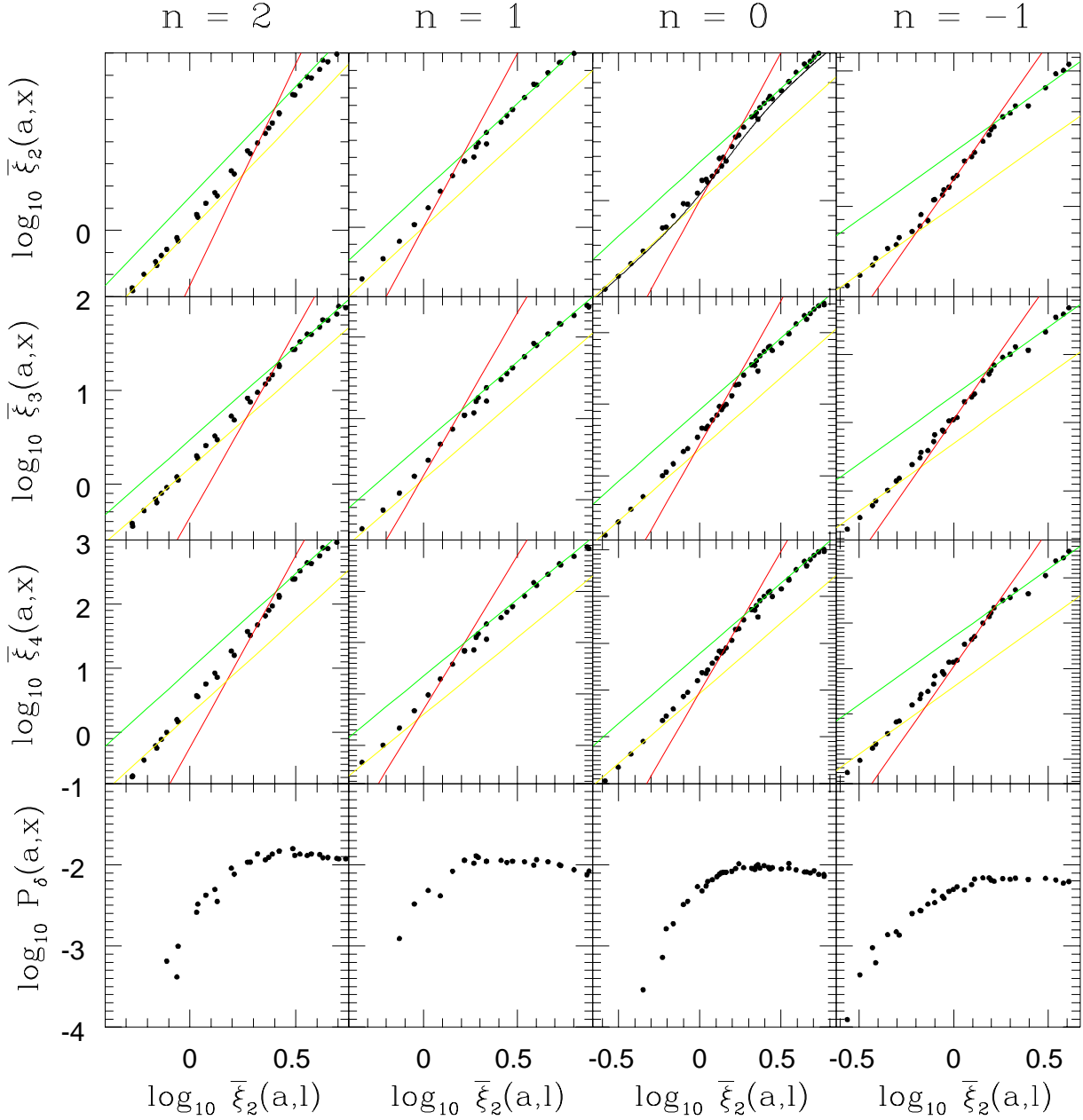}
}
\caption{The volume-averaged $N$--point correlation functions
$\bar \xi_N(x,a)$ are plotted against the linear
two-point function $\bar \xi_2(l,a)$ for various power law  
spectra. Straight lines in different
panels corresponds to slopes $( N - 1)$, $2(N-1)$ and $(N-1)$ 
for the perturbative, intermediate and highly nonlinear regimes, respectively.
Dots represent 2D N-Body simulation data. The lowest panel shows scaling 
in the pdf for $\delta = 5$.}
\end{figure*}

\section{Comparison with Simulation Results}
We have tested these ideas by examining the results of 2D N-body simulations
of gravitational clustering, based on the PM algorithm (cf. Beacom et al.
1991). We have run simulations of a variety of realisations of
initial power-law spectra, with $P(k) \propto k^{n}$ with random-phase
initial conditions generated in the standard manner. (Recall that
a power-law of index $n$ in 2D corresponds to a 3D spectral index of $n-1$.)

The variance at a fixed scale $x$ shows a nonlocal scaling, and it also
induces scaling in the volume average of higher-order correlations functions
due to hierarchal nature of N-point correlation functions.  
We can extract moments corresponding to different $x$ and $a$, 
$\bar \xi_N(x,a)$ 
from a series of different timesteps of the evolution of a single set
of initial data, in much the same way as was done by Jain, Mo \& White
(1995), and relate them to the the linearly extrapolated variance, $\xi_2(l,a)$, on the appropriate length scale $l = x( 1 + \bar \xi_2(x,a))^{1/2}$.
Determination of higher order correlations are non-trivial due to the presence
of final volume effect. We use  methods  based on factorial moments of the
counts-in-cells  to determine volume averages of higher order correlation 
functions; details of the correction procedure are explained by 
Munshi et al. (1997).

The results obtained by this approach can be used to populate a locus in the
plane of $\bar \xi_N(x,a)$ against $\bar \xi_2(l,a)$ as shown in Figure 1 for 
$N=2$,
$3$ and $4$. If scaling holds, the locus should be well defined
for results obtained from different evolutionary stages. 
Comparison of the theoretical 
slopes at different regimes of gravitational clustering with our 
simulations show that the scaling arguments presented in
Section 2 do indeed 
work fairly well for all the initial spectra considered, lending
further credence to the physical arguments from which they were developed.

It is also clear that agreement with theory is better for models with  
less power on small scales, the intermediate 
slope in  case of $n = 2$ spectrum shows pronounced deviation from its predicted value $2$. This is not altogether
surprising, as this model displays very strong clustering on small scales.
Since the scaling argument is based on spherical collapse, which might be
disrupted by strong sub-clustering, one might have anticipated a better match
of results for spectra with less power on small scales. A similar
breakdown might be anticipated for results obtained
from perturbation theory,
due to the  presence of large amounts of small scale 
power (Munshi et al. 1997).

Determination of higher- and higher-order correlations from simulation data 
becomes increasingly difficult due to their sensitivity to various spurious 
effects, such as the discrete nature and finite size of simulations. As 
described  in Section 2, the presence of nonlocal scaling in the 
two-point correlation functions results in a scaling of the entire 
pdf for scale free simulations.  Scaling behaviour in the pdf 
is much easier to study than its higher order moments. In Figure 1,
we present results for the pdf evaluated at an overdensity $\delta = 5 $. Studying
such nonlocal scaling 
might help developing a complete theory of evolution of pdf and its higher
order moments in different regimes (Colombi et al. 1996), without having
to worry about the errors in estimating the high-order correlations from
relatively small simulation volumes.

The other issue we sought to address is whether the stable clustering limit
is indeed reached during these simulations. However, the results displayed
in Figure 1 do not go very far into the strongly non-linear regime. The
extremely high resolution of our 2D simulations allows us to calculate
much higher variances than can be displayed on Figure 1 without
compressing the intermediate regime too much. In fact we find that, although the
early stages of the strongly non-linear regime do seem to be well described by
the stable clustering ansatz, the curves do show some deviations 
at very high values of $\bar \xi_2$. Such a behaviour might be expected if the
simulation were dominated by resolution problems, but we think this is
unlikely to be the case here, because the scale concerned is still
above the resolution length of the simulations. We also note that the
scaling observed has the same form for each evolutionary epoch in the
simulations. In the 3D study by Jain, Mo \& White (1995) it was found that
resolution effects resulted in a deviation from the stable clustering form
at different values of $\bar \xi_2 (x,a)$ for different evolutionary stages
(recall that the curves shown in Figure 1 are constructed by co-adding results
from different scales and simulation epochs). We do not find this to be the
case and consequently argue that the departure from stable clustering we see
is not a result of the same resolution problems they found.
Moreover, we find that
asymptotic power index for correlation functions at very high values of 
$\bar \xi_2$
becomes roughly 
independent of the slope of the 
initial power spectrum as has been suggested
by Padmanabhan (1996); see also Saslaw (1980). 
This would imply that the parameter $h$ can expressed 
as a function of initial power index $n$ according to
\begin{equation}
h = \frac{K}{n+2},
\end{equation}
where $K$ is a constant. Figure 2 shows the behaviour of the measured
$h$ as a function of $n+2$ for these simulations. It does indeed appear to 
be roughly satisfied although there is significant scatter in the plot. 
Notice that  $h=1$, independent of $n$, if stable clustering holds.
We feel that the results displayed in Figure 2 are unlikely to represent
a resolution effect, because the transition from the initial phase of stable
clustering to this later one is extremely smooth and noise-free. Other
claims to have detected stable clustering are much easier to interpret in terms
of the smaller dynamical range of available 3D simulations (Jain, Mo \& White 1995; 
Jain 1997). For example,
Jain (1997) shows how $h$ moves towards unity as $\bar \xi_2(x,a)$ increases but his
plots cut off before there is any evidence of a period where $h$ is
constant at a value of unity. In our simulations, $h$ carries on decreasing
below $h=1$ and shows no sign of asymptoteing to this value.
We have used the slope of scaling curves to measure the values of $h$,
other possibility include measuring directly the mean pair velocity. 
As mentioned in Jain (1997), the  mean pair velocity displays
a large scatter around
its mean value and a large number of realizations are needed to reduce
this dispersion. He used differential form of pair conservation equations to
evaluate $h$ from simulation data, equivalently one can use the 
scaling function directly in its integral form to find out the slope 
and hence the value of $h$.

\begin{figure}
\protect\centerline{
\epsfysize = 3.5truein
\epsfbox[20 146 587 714]
{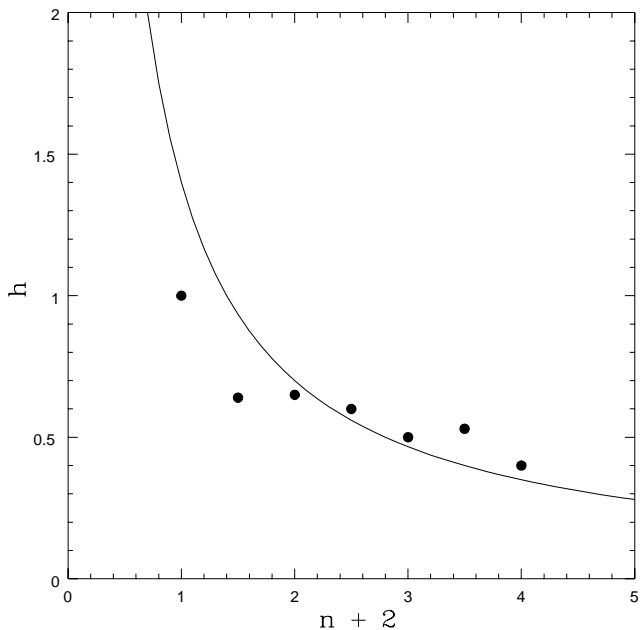}
}
\caption{The value of $h$ extracted from the simulations is 
plotted as a function of $n+2$. The solid curve
represents the relationship $h(n+2) = 1.4$, discussed in the text.  }
\end{figure}

\section{Conclusions}
We have addressed two main issues in this paper. The first is the extent to
which 2D gravitational clustering can be as well described by simple scaling
models as the 3D case appears to be. More 
importantly, we also wished to determine whether the quantitative
scaling behaviour in 2D is as expected on the basis of radial collapse
arguments. The answer to both of these questions seems to be ``yes'', 
suggesting that the scaling arguments used are  physically
well-motivated and provide a very good description of the simulation data.

The second issue we investigated was whether we could infer from the
behaviour of these simulations in the strongly non-linear regime that
a regime of stable clustering is reached. The answer to this appears to be
``no''. Although there is an initial strongly non-linear period of stable
clustering, there appears to be a significant departure from this behaviour
at very small scales.
 Of course, it is possible that at even higher clustering amplitudes
still, $h$ starts increasing and then settles down at a unit value. All we can say
is that there is no evidence from our study that this happens and consequently
that if stable clustering ever applies, it can only do so on exceedingly
small scales. 

We hope to investigate this second issue further in related studies. For example, 
a stability analysis of the coupled integro-differential equations 
which govern the behaviour of evolution of correlation functions around 
self-similar solution in the highly nonlinear regime can also
provide important insight into their asymptotic behaviour on small scales.
A recent study by Yano \& Gouda (1997) demonstrates that power law index
of the two point correlation function predicted by stable clustering ansatz
is not special in terms of stability of the solution. 
It is also worth mentioning that the
slope of the two-point correlation function can be related to 
the form of the halo  profile at very small radius. 
This can be done if we assume  that, at small
length scales, the dominant contribution to the correlation functions
comes from points within the same halo. 
A correlation function which becomes independent
of the initial power spectrum at very small scales would lead to
halo profiles independent of initial conditions at small radius.
Such a direct study of halo properties is,
in some sense, more fundamental than the scaling of the correlation
functions because while the correlation function can be calculated
if halo density profiles are known, the
converse is not necessarily the case.
There has been considerable interest in recent years devoted to this
question in 3D (Navarro et al. 1995; Cole \& Lacey 1996; 
Navarro et al. 1996; Syer \& White 1997). We intend to
complement these studies with a related investigation 
in 2D, which will again provide greater dynamical range with which 
to answer such questions.

\section*{Acknowledgments}
Dipak Munshi acknowledges support from PPARC under the QMW Astronomy 
Rolling Grant GR/K94133.
It is pleasure for Dipak Munshi to acknowledge Francis Bernardeau and Richard
Schaeffer for many useful discussions regarding evaluation of the
$S_N$ parameters.  Peter Coles receives a PPARC Advanced Research Fellowship. 
We are grateful for support under the NSF-EPSCoR program, as well as the 
Visiting Professorship and General Funds of the University of Kansas.

\end{document}